\documentclass[11pt]{article}

\usepackage{moriond}
\usepackage{graphicx}
\usepackage{cite}

\newcommand{\epem}              {\ensuremath{\mathrm{e^+e^-}}}
\newcommand{\as}                {\ensuremath{\alpha_\mathrm{S}}}
\newcommand{\mz}                {\ensuremath{M_{\mathrm{Z^0}}}}
\newcommand{\asmz}              {\ensuremath{\as(\mz)}}
\newcommand{\oa}                {\ensuremath{\mathcal{O}(\as)}}
\newcommand{\oaa}               {\ensuremath{\mathcal{O}(\as^2)}}
\newcommand{\oaaa}              {\ensuremath{\mathcal{O}(\as^3)}}
\newcommand{\rs}                {\ensuremath{\sqrt{s}}}

\newcommand{\bbbar}             {\ensuremath{\mathrm{b\bar{b}}}}
\newcommand{\ycut}              {\ensuremath{y_{\mathrm{cut}}}}
\newcommand{\ymax}              {\ensuremath{y_{\mathrm{max}}}}
\newcommand{\stat}              {\ensuremath{\mathrm{(stat.)}}}

\newcommand{\expt}              {\ensuremath{\mathrm{(exp.)}}}
\newcommand{\had}               {\ensuremath{\mathrm{(had.)}}}
\newcommand{\theo}              {\ensuremath{\mathrm{(theo.)}}}
\newcommand{\thr}               {\ensuremath{1-T}}
\newcommand{\mh}                {\ensuremath{M_\mathrm{H}}}
\newcommand{\bt}                {\ensuremath{B_\mathrm{T}}}
\newcommand{\bw}                {\ensuremath{B_\mathrm{W}}}
\newcommand{\cp}                {\ensuremath{C}}
\newcommand{\ytwothree}        {\ensuremath{y_{23}}}
\newcommand{\momone}[1]         {\mbox{\ensuremath{\langle#1\rangle}}}
\newcommand{\momn}[2]           {\mbox{\ensuremath{\langle#1^{#2}\rangle}}}
\newcommand{\dd}                {\ensuremath{\mathrm{d}}}

\begin{document}


\title{ New QCD tests with old JADE data }

\author{S. Kluth}

\address{Max-Plank-Institut f\"ur Physik, F\"ohringer Ring 6, D-80805,
Germany \\
E-mail: skluth@mppmu.mpg.de }

\maketitle\abstracts{ Data from \epem\ annihilation into
  hadrons collected by the JADE experiment at centre-of-mass energies
  between 14 and 44~GeV were used to study the 4-jet rate using the
  Durham algorithm as well as the first five moments of event shape
  observables.  The data were compared with NLO QCD predictions,
  augmented by resummed NLLA calculations for the 4-jet rate, in order
  to extract values of the strong coupling constant \as.  The
  preliminary results are $\asmz=0.1169\pm0.0026$ (4-jet rate) and
  $\asmz=0.1286\pm0.0072$ (moments) consistent with the world average
  value.  For some of the higher moments systematic deficiencies of
  the QCD predictions are observed.  }

\section{ Introduction }

The production of hadrons in \epem\ annihilation allows precise tests
of the gauge theory of strong interactions, Quantum Chromodynamics
(QCD).  In this paper recent and preliminary analyses of JADE data
using the 4-jet rate based on the Durham algorithm~\cite{durham} and
using the first five moments of event shape observables are
presented~\cite{jadenote146,jadenote147}.  

The data used in our analyses were collected at centre-of-mass (cms)
energies $\rs=14.0$, 22.0, 34.6, 35.0, 38.3 and 43.8~GeV between 1981
and 1986 with the JADE detector~\cite{naroska87}.  The data samples
consist of ${\cal O}(1000)$ events at $\rs=14.0$, 22.0, 38.3 and 43.8~GeV
while at $\rs=34.6$ (35.0) GeV about 14000 (21000) events are used.

The software employed to perform the analyses includes the original
JADE detector simulation, event reconstruction and event display
programs, see~\cite{jadenote146,jadenote147} for details.  It is
possible to use recent Monte Carlo event generators such as PYTHIA,
HERWIG or ARIADNE to generate simulated events, to pass these through
the JADE detector simulation and to reconstruct them in essentially
the same way as the data.  The event generators were used with
parameter settings obtained by OPAL after adjusting to LEP~1 data.
The original data are only available after a further step of data
reduction has been performed, resulting in information about 4-vectors
of reconstructed particles and some quantities for event selection.

The selection of well reconstructed tracks from the tracking detectors
and clusters from the electromagnetic calorimeter as well as the
selection of hadronic events follows previous analyses.  Most
importantly requirements on visible energy and momentum, balance of
momentum along the beam direction and track multiplicity suppress
events from two-photon interactions, $\tau$ production and other
backgrounds~\cite{naroska87,jadenewas,OPALPR299,movilla02b}.

The contribution of $\epem\rightarrow\bbbar$ events is subtracted from
the data using simulated events.  Then the data are corrected for the
effects of detector resolution and acceptance of selection cuts by
correction factors determined from simulated events before and after
the JADE detector simulation.

Experimental systematic uncertainties of the analyses include
variation of the event selection cuts, comparison of data sets
resulting from different versions of the JADE event reconstruction
program and comparing experimental corrections derived from different
event generators.

Additional systematic uncertainties are studied for the extraction of
values of the strong coupling constant \as.  Hadronisation
uncertainties are evaluated by using different Monte Carlo generators
to compute the hadronisation corrections.  Theoretical systematic
uncertainties are found by changing the renormalisation scale $\mu$ of
the QCD predictions from $\mu=\rs$ to $\mu=\rs/2$ and $\mu=2\rs$.

\section{ 4-Jet Rate }
\label{sec_r4}

\begin{figure}[htb!]
\begin{tabular}{cc}
\includegraphics[width=0.45\columnwidth]{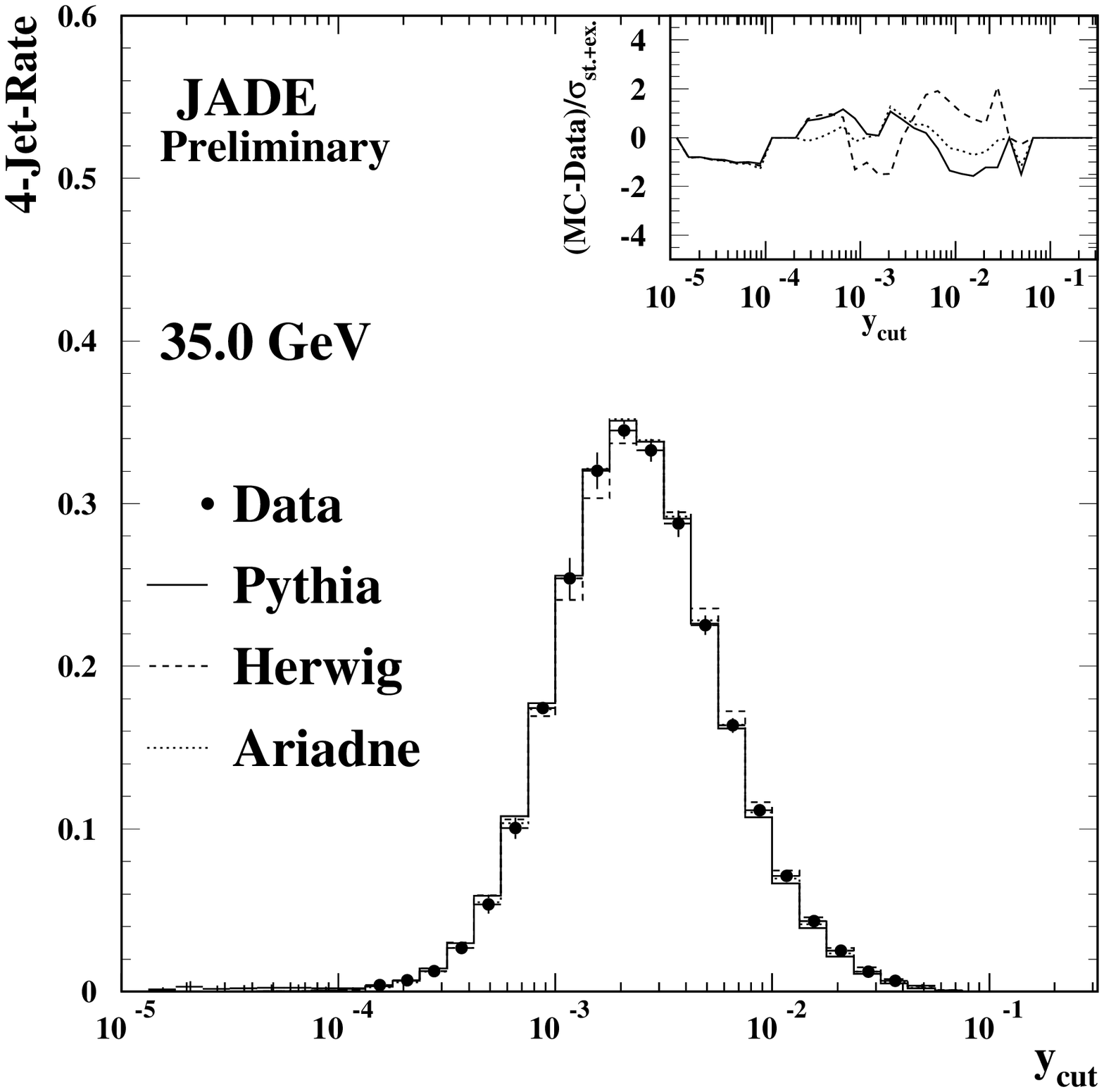} &
\includegraphics[width=0.45\columnwidth]{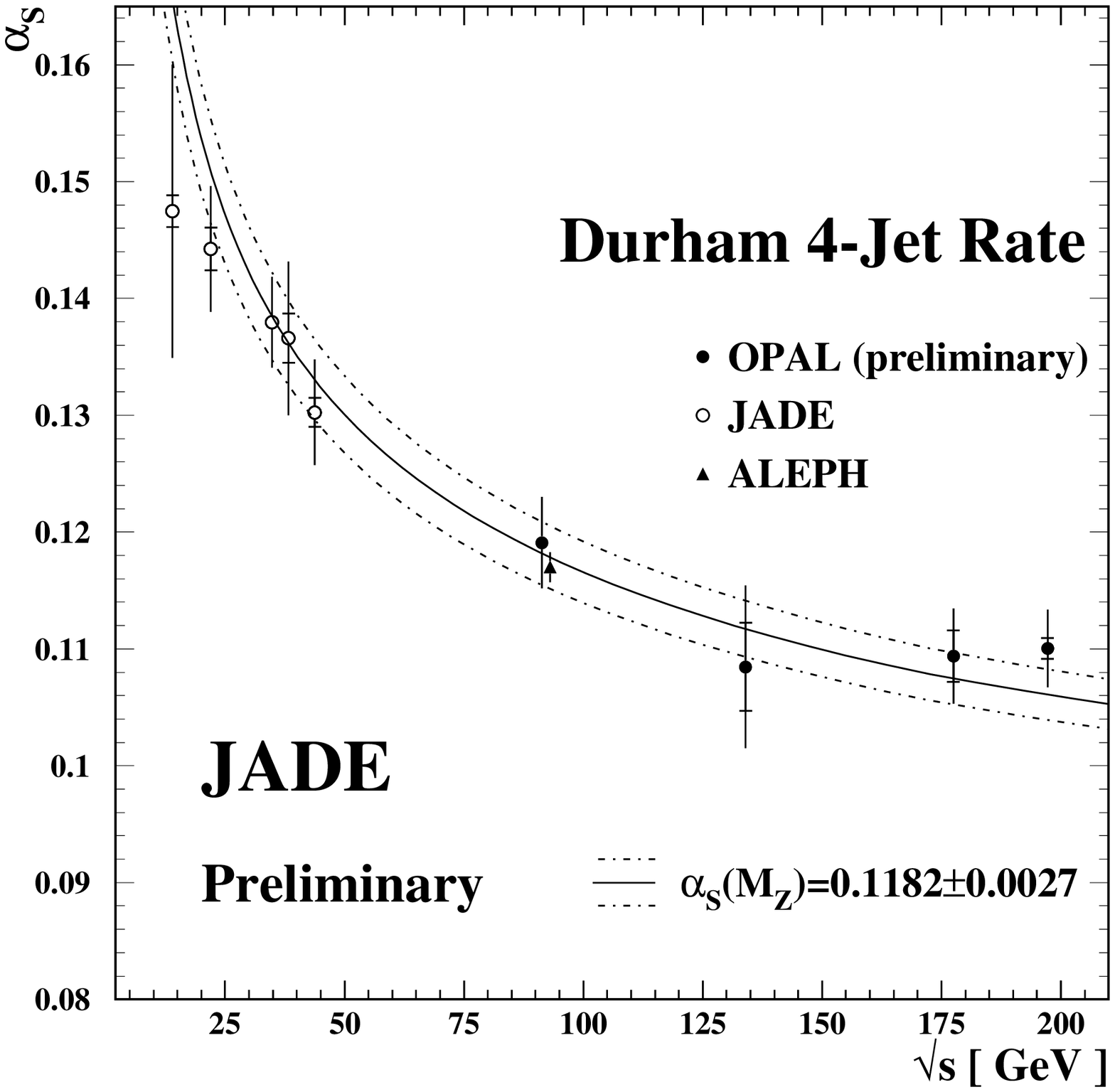} \\
\end{tabular}
\caption{ (left) The figure shows the corrected data for the 4-jet 
  rate at $\rs=35$~GeV compared with predictions from Monte Carlo
  models.  The error bars represent statistical and experimental
  errors added in quadrature~\cite{jadenote146}.  (right) The figure
  shows results for \as\ from the 4-jet rate as a function of \rs.
  The error bars show the statistical and total uncertainties.  The
  full and dash-dotted lines indicate the current world average value
  of \asmz~\cite{bethke04}.  The results at $\rs=34.6$ and 35~GeV have
  been combined for clarity.  Results from OPAL and ALEPH are shown as
  well~\cite{jadenote146}. }
\label{fig_r4}
\end{figure}

Jets are reconstructed in the hadronic events using the Durham jet
clustering algorithm~\cite{durham}.  Figure~\ref{fig_r4} (left) shows the
data for the 4-jet rate corrected for experimental effects as a
function of \ycut\ measured at $\rs=35$~GeV.  The data are compared to
predictions from the event generators PYTHIA, HERWIG and ARIADNE.  We
find good agreement between the data and the predictions within the
statistical and experimental errors and conclude that the models can
be used to correct for hadronisation effects.

The QCD predictions are next-to-leading-order (NLO), i.e.\ \oaa\ in
leading-order (LO) with \oaaa\ radiative corrections, combined with
resummed next-to-leading-logarithm (NLLA) calculations~\cite{nagy98b}.
We find good agreement between data and theory for large values of
$\ycut={\cal O}(10^{-2})$ where mostly 3- and 4-jet events are found
and the theory considering at most five partons is expected to be
reliable.  The resulting values of \as\ are shown in
figure~\ref{fig_r4} (right).

The values for \asmz\ from the data at $\rs=22.0$, 34.6, 35.0,
38.3 and 43.8~GeV are combined taking into account correlations
between experimental, hadronisation and theoretical systematic
uncertainties.  The fits at $\rs=14.0$~GeV have large experimental and
hadronisation uncertainties and are therefore excluded from the
average.  The result is
$\asmz=0.1169\pm0.0004\stat\pm0.0012\expt\pm0.0021\had\pm0.0007\theo$,
$\asmz=0.1169\pm0.0026$ (total error), consistent with the current
world average $\asmz=0.1182\pm0.0027$~\cite{bethke04}.  As shown 
in~\cite{nagy98b} the theoretical uncertainty rises with small \ycut\
but is small within the fitranges.

\section{ Moments of Event Shape Observables }

The first five moments of the distributions of the event shape
observables \thr, \cp, \bt, \bw, \ytwothree\ and \mh\ are calculated
according to $\momn{y}{n}=\int_0^{\ymax} y^n\, 1/\sigma\,
\dd\sigma/\dd y\, \dd y'$, where $y$ denotes one of the observables,
\ymax\ is the kinematically allowed upper limit of the observable and
$n=1,\ldots,5$.

The calculation of perturbative QCD predictions in NLO (\oa\ in LO
with \oaa\ radiative corrections) involve a full integration over
phase space.  This analysis is thus complementary to tests
of the theory using the differential distributions which are commonly
only compared with data in restricted regions, where the theory is
able to describe the data well, see e.g.~\cite{jadenewas}.

\begin{figure}[htb!]
\begin{tabular}{cc}
\includegraphics[width=0.35\columnwidth]{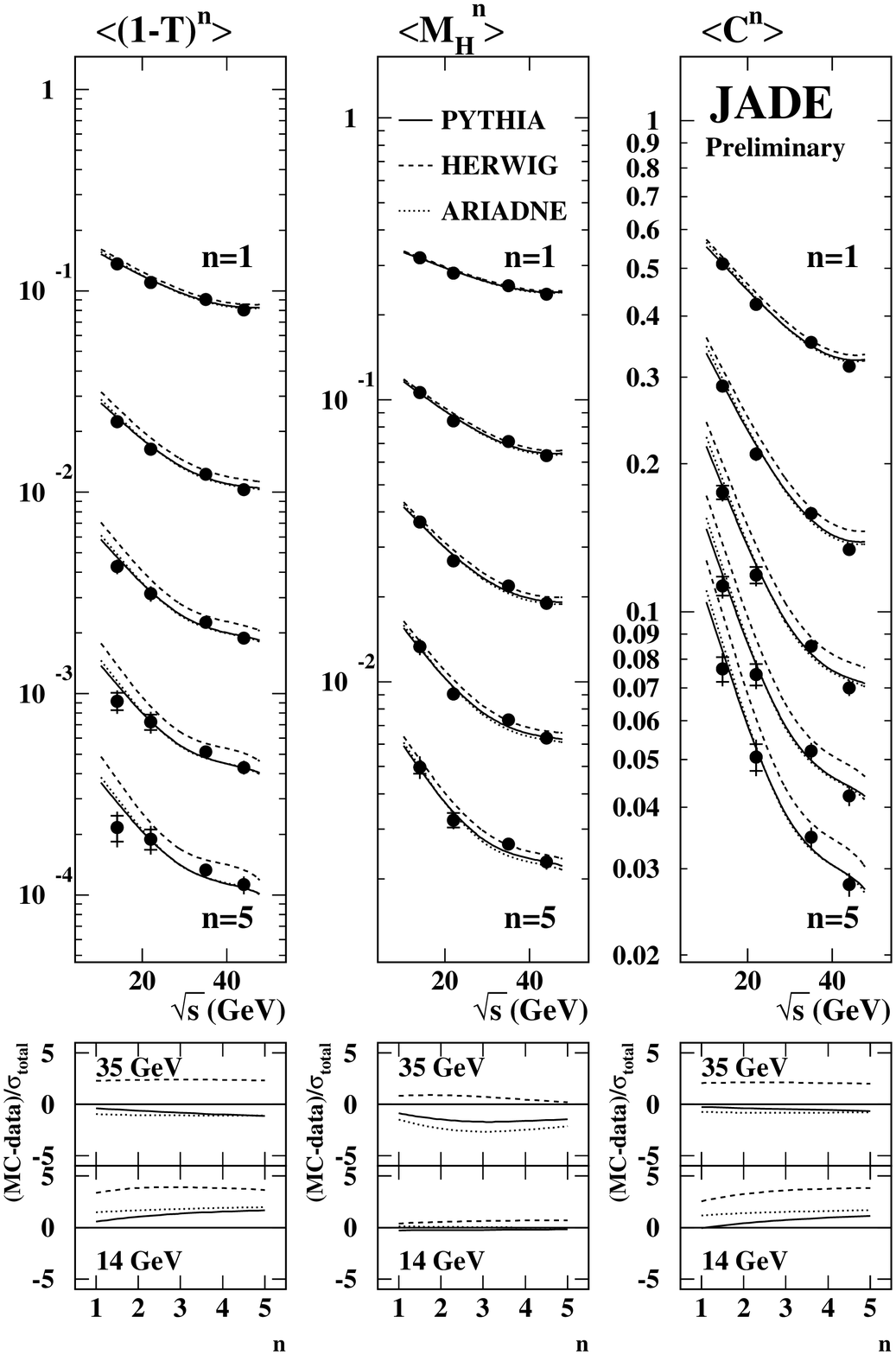} &
\includegraphics[width=0.55\columnwidth]{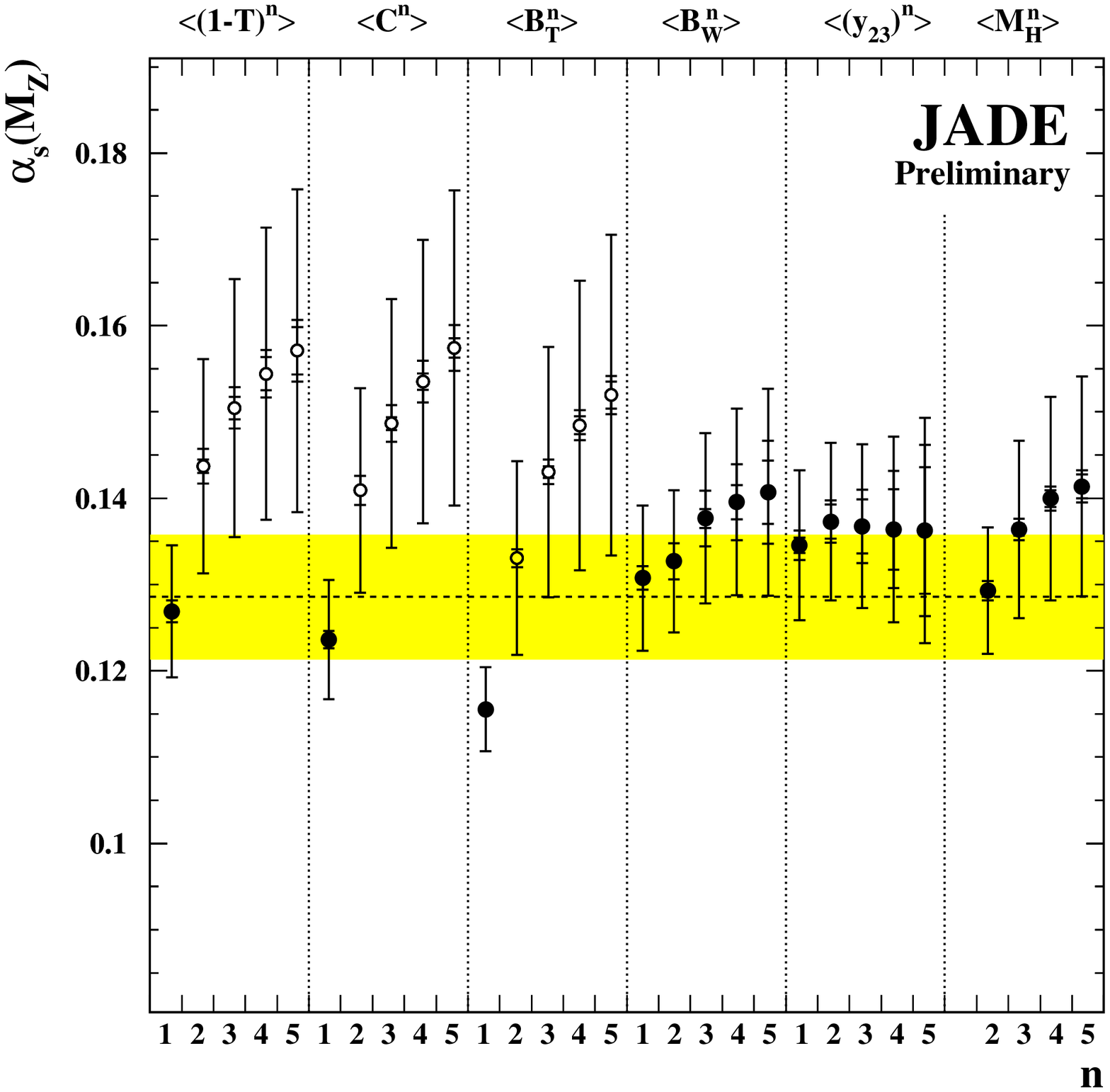} \\
\end{tabular}
\caption{ (left) The corrected data for the first five moments of 
  the observables \thr, \mh\ and \cp\ are presented with error bars
  showing statistical and experimental error added in quadrature. The
  lines indicate the predictions of Monte Carlo models.  The lower
  panels show the differences between data and models at $\rs=14$ and
  35~GeV, divided by the errors~\cite{jadenote147}.  (right)
  Measurements of \asmz\ using fits to moments of six event shape
  observables are shown.  The inner error bars represent statistical
  errors, the middle error bars include experimental errors and the
  outer error bars show the total errors.  The dotted line indicates
  the weighted average described in the text; only the measurements
  indicated by solid symbols were used for this
  purpose~\cite{jadenote147}.  }
\label{fig_mom}
\end{figure}

Figure~\ref{fig_mom} (left) presents the data for the first five
moments of \thr, \mh\ and \cp\ corrected for experimental effects
compared with predictions by the same event generators as in
section~\ref{sec_r4}.  There is generally good agreement between data
and model predictions; HERWIG is seen to describe the data somewhat
less well than PYHTIA or ARIADNE.  We will use the models to derive
hadronisation corrections in order to compare the data with
predictions from perturbative QCD.

We fitted the QCD predictions corrected for hadronisation to the data
for a given observable and moment $n=1,\ldots,5$ individually with
\asmz\ as the only free parameter.  The results for \asmz\ are
summarised in figure~\ref{fig_mom} (right).  The fit to \momone{\mh}
did not converge and therefore no result is shown.  We observe that
the values of \asmz\ increase with $n$ for the observables
\momn{(1-T)}{n}, \momn{\cp}{n} and \momn{\bt}{n}, while for the other
observables \momn{\bw}{n}, \momn{(\ytwothree)}{n} and \momn{\mh}{n},
$n=2,\ldots,5$, the results are fairly stable.

We evaluated the ratio $K$ of NLO and LO coefficients for the six
observables used in our fits and found a clear correlation between the
steeply increasing values of \asmz\ and increasing values of $K$ with
$n$ for \momn{(1-T)}{n}, \momn{\cp}{n} and \momn{\bt}{n}.  The other
observables \momn{\bw}{n}, \momn{(\ytwothree)}{n} and \momn{\mh}{n},
$n=2,\ldots,5$, have fairly constant values of $K$ and correspondingly
stable results for \asmz.  We also noted that \momone{\mh} has a large
and negative value of $K$ which is the cause that the fit did not
converge.

In order to find a combined value of \asmz\ we considered only those
results for which the NLO term is less than half the LO term (i.e.\ 
$|K\as/2\pi|<0.5$), namely \momone{\thr}, \momone{\cp}, \momone{\bt},
\momn{\bw}{n} and \momn{(\ytwothree)}{n}, $n=1,\ldots,5$ and
\momn{\mh}{n}, $n=2,\ldots,5$; i.e.\ results from 17 observables in
total.  The purpose of this requirement was to select observables with
an apparently converging perturbative prediction.  Correlations
between statistical, experimental, hadronisation and theoretical
uncertainties were considered when forming the average.  The result is
$\asmz=0.1286\pm0.0007\stat\pm0.0011\expt\pm0.0022\had\pm0.0068\theo$,
$\asmz=0.1286\pm0.0072$ (total error), above but still consistent with
the world average value.  It has been observed previously in
comparisons of distributions of event shape observables with NLO QCD
predictions with renormalisation scale $\mu=\rs$ that fitted values of
\asmz\ tend to be large, see e.g.~\cite{OPALPR075}.

\section{ Summary }

We have presented preliminary results of measurements of the 4-jet
rate based on the Durham algorithm and the first five moments of event
shape observables using JADE data at $\rs=14.0$ to 43.8~GeV.  The
predictions of the Monte Carlo models PYTHIA, HERWIG and ARIADNE tuned
by OPAL to LEP~1 data were found to be in reasonable agreement with
the data.  The data have also been used to extract measurements of the
strong coupling constant \asmz\ with the results $\asmz=0.1169\pm0.0026$
(4-jet rate) and $\asmz=0.1286\pm0.0072$ (moments) in agreement with
the world average value.  The higher moments of \thr, \cp\ and \bt\ are
observed to yield systematically larger values of \asmz.

\end{document}